\documentstyle[12pt]{article}

\newcommand{\be}{\begin{equation}}
\newcommand{\ee}{\end{equation}}
\newcommand{\om}{\omega}
\newcommand{\Om}{\Omega}
\newcommand{\vp}{\varphi}
\newcommand{\dgr}{\dagger}
\newcommand{\ra}{\rightarrow}
\newcommand{\vr}{{\bf r}}
\newcommand{\vk}{{\bf k}}
\newcommand{\prt}{\partial}
\newcommand{\dlt}{\delta}
\newcommand{\sgm}{\sigma}
\newcommand{\Dlt}{\Delta}
\newcommand{\al}{\alpha}  

\begin{document}

\begin{center}

{\Large{\bf Matrix Order Indices in Statistical Mechanics} \\ [5mm]
V.I. Yukalov } \\ [3mm]

{\it Research Center for Optics and Photonics \\

Instituto de Fisica de S\~ao Carlos, Universidade de S\~ao Paulo \\

Caixa Postal 369, S\~ao Carlos, S\~ao Paulo 13560-970, Brazil \\

and \\

Bogolubov Laboratory of Theoretical Physics \\

Joint Institute for Nuclear Research, Dubna 141980, Russia}

\end{center}

\vskip 2cm

\begin{abstract}

A new notion is introduced of matrix order indices which relate the 
matrix norm and its trace. These indices can be defined for any given 
matrix. They are especially important for matrices describing many-body 
systems, equilibrium as well as nonequilibrium, for which the indices 
present a quantitative measure of the level of ordering. They 
characterize not only the long-range order, but also mid-range order. 
In the latter case, when order parameters do not exist, the matrix 
indices are well defined, providing an explicit classification of 
various mid-range orders. The matrix order indices are suitable for 
describing phase transitions with both off-diagonal and diagonal order. 
Contrary to order parameters whose correct definition requires the 
thermodynamic limit, the matrix indices do not necessarily need the 
latter. Because of this, such indices can distinguish between different 
phases of finite systems, thus, allowing for the classification of 
crossover phase transitions.

\end{abstract}

\vskip 1cm


\vskip 1cm


\newpage

\section{Introduction}

Matrices are common objects in quantum physics, where all observable 
quantities are related with matrix representations of operators. 
In statistical mechanics, a special role is played by reduced density 
matrices [1-3]. The eigenvalues of the latter are known to be associated 
with a possible existence of long-range order called off-diagonal long-range
order [4--6]. This linkage between the eigenvalues and the order has 
been mathematically formalized by means of order indices for reduced 
density matrices [7--9]. These indices turned out to be a convenient 
tool for characterizing the arising order related to Bose-Einstein 
condensation and superconductivity [3]. There are, however, several 
limitations not permitting one to employ these indices for describing 
other types of order. Among the most important problems, the following 
are to be solved: (i) How the appearance of magnetic order could be 
noticed in the behaviour of the indices? (ii) How to describe phase 
transitions associated with diagonal order, such as crystallization? 
(iii) How to characterize crossover-type phase transitions occurring 
in finite systems?

In the present paper a new notion of matrix order indices is introduced. 
This notion is an essential generalization of the order indices for 
reduced density matrices [7--9]. Contrary to the latter, the matrix 
indices are defined for any given matrix of arbitrary nature. Because 
of this generality, the matrix indices are suitable for characterizing 
any types of phase transitions, accompanied by any
kinds of order, diagonal as well as off-diagonal, long-range as 
well as mid-range, in infinite as well as in finite systems. In section 2, 
a general mathematical definition of the matrix order indices is given. 
These are the quantities linking the norm and the trace of a given matrix 
whose nature can be arbitrary. The possibility of using different norms 
is discussed. In section 3, the matrix indices of self-adjoint semipositive
 matrices are considered, because of the special role played by such 
matrices in physical applications. The convenience of employing the 
introduced matrix indices for quantitatively comparing various kinds 
of mid-range orders is illustrated in section 4. In section 5, 
Bose-Einstein condensation in confined systems is analysed, which 
is important for describing the condensation of trapped atoms. 
Section 6 demonstrates how the matrix indices must be defined in order 
to treat magnetic phase transitions. Finally, in section 7, it is shown 
how the phase transitions, such as crystallization, characterized by 
diagonal order, can be described by an appropriate definition of the 
matrix order indices.

\section{Matrix order indices}

Let us have a matrix $\hat M$ of arbitrary nature, with a norm 
$||\hat M||$ and trace ${\rm Tr}\hat M$. The {\it matrix order index} 
is defined as
\be
\label{1}
\om_M \equiv \frac{\ln||\hat M||}{\ln|{\rm Tr}\hat M|} \; .
\ee
We shall say that among two given matrices that one is better ordered 
whose order index is larger. As follows from definition (1), for norm-one 
and traceless matrices, one has $\om_M=0$.

The value of the matrix index (1) depends on the type of the norm 
$||\hat M||$ used. Assume that a matrix $\hat M$ acts on a normed space 
$\{\vp\}$ of functions having a vector norm $|\vp|$. The matrix norm, 
associated with a vector norm $|\vp|$, is given by
$$
||\hat M|| \equiv \sup_\vp \; \frac{|\hat M\vp|}{|\vp|} \; .
$$
Here and in what follows, it is implied that $|\vp|\neq 0$. If $\{\vp\}$ 
is a Hilbert space, with a scalar product $(\vp,\vp)$, the latter naturally 
generates the Hermitian vector norm $|\vp|=\sqrt{(\vp,\vp)}$. Then, for a 
matrix $\hat M$, the {\it Hermitian norm} is
\be
\label{2}
||\hat M|| \equiv \sup_\vp \; \frac{(\vp,\hat M^+ \hat M\vp)^{1/2}}{|\vp|} \; ,
\ee
where $\hat M^+$ is a Hermitian conjugated matrix. Supposing that the matrix 
$\hat M$ possesses a spectrm $\{\mu_n\}$ of eigenvalues, one has the 
{\it spectral norm}
\be
\label{3}
||\hat M|| = \sup_n |\mu_n|\; .
\ee
More generally, the spectral norm is defined as the square root of the 
largest eigenvalue of the matrix $\hat M^+ M$. When $\hat M=\hat M^+$ is 
a self-adjoint and semipositive matrix, then its eigenvalues are semipositive,
$\mu_n\geq 0$. In that case, the spectral norm becomes
\be
\label{4}
||\hat M|| = \sup_\vp \; \frac{(\vp,\hat M\vp)}{|\vp|} = 
\sup_n \mu_n \; .
\ee

Among other matrix norms, for physical applications one may employ the 
{\it trace norm}
\be
\label{5}
||\hat M||' \equiv \left ( {\rm Tr} \hat M^+ \hat M\right )^{1/2} \; ,
\ee
which is also called the Euclidean norm. Definition (5) may be convenient 
for calculations, since the trace does not depend on a matrix representation. 
With a spectrum $\{\mu_n\}$ of $\hat M$, the trace norm (5) reads
$$
||\hat M||' = \left ( \sum_n |\mu_n|^2\right )^{1/2} \; .
$$
As follows from the above definitions, $||\hat M||\leq ||\hat M||'$, hence 
one has to distinguish between the associated indices $\om_M$ and $\om_M'$, 
for which $\om_M\leq\om_M'$.

\section{Self-adjoint semipositive matrices}

Self-adjoint matrices play a special role in physical applications, because 
of which they need to be studied more attentively. Let ${\cal X}=\{ x\}$ 
be an arithmetic space of physical coordinates and let $t\in[0,\infty)$ 
denote time. The set $x$ may include Cartesian coordinates, spin, isospin, 
and other quantum variables. Assume that an algebra of local observables 
${\cal A}=\{ A(x,t)\}$ and an algebra of field operators 
$\Psi=\{\psi(x,t),\psi^\dgr(x,t)\}$ are given on a Fock space ${\cal F}$. 
The direct sum $\tilde{\cal A}\equiv{\cal A} \oplus\Psi$ denotes an 
{\it extended local algebra}.

Consider a system of $p$ objects that, for concreteness, will be called 
particles. The set $x^p\equiv\{ x_1,x_2,\ldots,x_p\}$ of particle 
coordinates pertains to the space ${\cal X}^p\equiv{\cal X}_1\times
{\cal X}_2\times\ldots \times{\cal X}_p$, which is a $p$-fold tenzor 
product. In a Hilbert space of complex functions $\vp_p(x^p)$, the scalar 
product is given by the integral
$$
(\vp_p,\vp_p') \equiv \int \vp_p^*(x^p)\; \vp_p'(x^p)\; dx^p \; ,
$$
where $dx^p \equiv dx_1dx_2\ldots dx_p$. In general, $p=1,2,\ldots$.

Introduce the function
\be
\label{6}
D_p(A,x^p,\overline x^p,t) \equiv {\rm Tr}_{\cal F} A(x_1)\ldots A(x_p)
\hat\rho(t) A^+(\overline x_p)\ldots A^+(\overline x_1) \; ,
\ee
where the trace is taken over the Fock space ${\cal F}$, 
$A(x)\equiv A(x,0)$ is a member of the extended local algebra 
$\tilde{\cal A}$, and $\hat\rho(t)$ is a statistical operator. It is 
admissible to treat the function (6) as an element of the matrix
\be
\label{7}
\hat D_p(A,t) \equiv [ D_p(A,x^p,\overline x^p,t)]
\ee
with respect to the variables $x^p$. By construction (6), the matrix (7) 
is self-adjoint and positive semidefinite. The action of the matrix (7) 
on a column $\vp_p\equiv[\vp_p(x^p)]$ gives the column 
$$
\hat D_p(A,t)\vp_p = \left [ 
\int D_p(A,x^p,\overline x^p,t)\; \vp_p(\overline x^p)\;
d\overline x^p \right ] \; .
$$
A linear envelope of the vectors $\vp_p$ forms a Hilbert space 
${\cal H}_p={\cal L}\{ \vp_p\}$ with a scalar product $\vp_p^+\vp_p'\equiv
(\vp_p,\vp_p')$. Each matrix (7) acts on ${\cal H}_p$. The spectral norm 
for a self-adjoint semipositive matrix can be written as
\be
\label{8}
||\hat D_p(A,t)|| \equiv \sup_{\vp_p} \; 
\frac{\vp_p^+\hat D_p(A,t)\vp_p}{\vp_p^+\vp_p} \; .
\ee
The trace over the variables $x^p$ is
\be
{\rm Tr}\; \hat D_p(A,t) \equiv \int D_p(A,x^p,x^p,t)\;
dx^p \; .
\ee
According to Eq. (1), one may introduce the matrix indices
\be
\label{10}
\om_p(A,t) \equiv 
\frac{\ln||\hat D_p(A,t)||}{\ln|{\rm Tr}\; \hat D_p(A,t)|} \; .
\ee
For a semipositive matrix, one has
$$
||\hat D_p(A,t)|| \leq {\rm Tr}\; \hat D_p(A,t) \; ,
$$
because of which
\be
\label{11}
\om_p(A,t) \leq 1 \; .
\ee
Similarly, employing the trace norm (5), one obtains the matrix indices 
$\om_p'(A,t)$ for which inequality (11) is also valid.

For constructing a matrix (7), one may choose any representative of the 
extended local algebra $\tilde{\cal A}$. In particular, opting for the 
field operators $\psi(x,t)$, we come to the reduced density matrices
\be
\label{12}
\hat\rho_p(t) \equiv \left [ \rho_p(x^p,\overline x^p,t)\right ]
\ee
with the elements
\be
\label{13}
\rho_p(x^p,\overline x^p,t) \equiv D_p(\psi,x^p,\overline x^p,t) =
{\rm Tr}_{\cal F} \; \psi(x_1)\ldots \psi(x_p) \hat\rho(t)
\psi^\dgr(\overline x_p) \ldots \psi^\dgr(\overline x_1) \; .
\ee
The spectral norm for the $p$-particle density matrix (12) is
\be
\label{14}
||\hat\rho_p(t)|| = \sup_n \rho_{pn}(t) \; ,
\ee
with $\{\rho_{pn}(t)\}$ being the related spectrum.

The number of particles, $N$, in a statistical system is assumed to be 
large. Then, in the trace
$$
{\rm Tr}\; \hat\rho_p(t) = \frac{N!}{(N-p)!} \qquad
(p=1,2,\ldots,N)
$$
one may use the Stirling formula $N!\simeq\sqrt{2\pi N} N^N e^{-N}$, which, 
under $p$ fixed and $N\gg 1$, yields
$$
\frac{N!}{(N-p)!} \simeq \frac{N^p}{e^p} \; .
$$
Hence, the matrix index (10) becomes
\be
\label{15}
\om_p(\psi,t) \simeq \frac{\ln||\hat\rho_p(t)||}{p\ln N} 
\ee
for $N\gg 1$.

For equilibrium statistical systems, one often considers the thermodynamic 
limit
\be
\label{16}
N\ra \infty\; , \qquad V\ra \infty\; , \qquad
\frac{N}{V}\ra \rho\; ,
\ee
in which $N$ is the number of particles in the volume $V$ and the limiting 
density $\rho$ is finite, $0<\rho<\infty$. In equilibrium, the matrix index 
(15) does not depend on time, becoming 
\be
\label{17}
\om_p(\psi) \simeq \frac{\ln||\hat\rho_p||}{p\ln N} 
\ee
as $N\gg 1$. Note that for calculating the index (17) it is not compulsory 
to take the actual thermodynamic limit (16) by setting $N\ra\infty$, but 
Eq. (17) can be used for finite systems, provided that $N\gg 1$.

From the properties of the reduced density matrices [3], we know that for 
bosons $||\hat\rho_p||\leq N^p\cdot const$, hence $\om_p(\psi)\leq 1$, in 
agreement with the general inequality (11). The equality
\be
\label{18}
\om_p(\psi) = 1 \qquad (boson \; long-range \; order)
\ee
corresponds to long-range order in a bosonic system, as that occurring at 
Bose-Einstein condensation of dilute gases.

Reduced density matrices for fermionic systems [3] satisfy the 
inequalities
$$
||\hat\rho_{2p}|| \leq N^p \cdot const\; , \qquad
||\hat\rho_{2p-1}|| \leq N^{p-1}\cdot const \; ,
$$
because of which for the matrix indices we have
\be
\label{19}
\om_{2p}(\psi) \leq \frac{1}{2} \; , \qquad \om_{2p-1}(\psi) \leq
\frac{p-1}{2p-1} \; .
\ee
The limiting case
\be
\label{20}
\om_{2p}(\psi) = \frac{1}{2} \qquad (even \; long-range \; order)
\ee
happens when even long-range order develops in a fermionic system, like 
that arising in superconductors. When Eq. (20) is valid, then
$$
\om_{2p-1}(\psi) = \frac{p-1}{2p-1} \qquad (p=1,2,\ldots,N) \; .
$$
For large $p$, this tends to the value (20), since
$$
\om_{2p-1}(\psi) \simeq \frac{1}{2} \; - \; \frac{1}{4p} \qquad
(p\gg 1) \; .
$$
Even long-range order may also happen for bosonic systems.

The matrix indices, introduced above, are different from the indices for 
reduced density matrices, defined in the earlier papers [7--9]. The main 
difference is that, in the case of long-range order, the matrix indices 
$\om_p(\psi)$ or $\om_{2p}(\psi)$ do not depend on $p$, displaying a kind 
of resonance, as follows from Eqs. (18) and (20).

\section{Finite momentum condensation}

In order to explicitly illustrate the calculation of the matrix order 
indices, for both long-range and mid-range orders, let us consider a Bose 
system with finite momentum condensate  [10--12]. Such a condensate 
consists of particles having a finite modulus of momentum $k_0=|\vk_0|$, 
with random directions of the vector $\vk_0$. The related condensate 
density matrix reads 
\be
\label{21}
\rho_0(\vr,\vr') =\frac{\rho_0}{\int d\Om(\vk_0)} \;
\int e^{i\vk_0\cdot(\vr-\vr')} \; d\Om(\vk_0) \; ,
\ee
where $\rho_0\equiv N_0/V$ is the condensate density and $\Om(\vk_0)$ is a 
spherical angle associated with $\vk_0$. The single-particle density matrix 
tends, as the distance between the points $\vr$ and $\vr'$ increases, to 
the condensate matrix (21),
\be
\label{22}
\rho_1(\vr,\vr') \simeq \rho_0(\vr,\vr') \qquad 
(|\vr -\vr'| \ra \infty ) \; .
\ee
In a particuar case of zero momentum $\vk_0=0$, Eq. (21) reduces to the 
constant density
$$
\rho_0(\vr,\vr') = \rho_0 \qquad (\vk_0\equiv 0)
$$
corresponding to the usual Bose-Einstein condensate with the order index
(18). But for the finite-momentum condensate, the situation is more 
elaborate and depends on the space dimensionality.

\subsection{Three-dimensional case}

In the three-dimensional space, the condensate matrix (21) reads
\be
\label{23}
\rho_0(\vr,\vr') = \rho_0\; 
\frac{\sin k_0|\vr-\vr'|}{k_0 |\vr -\vr'|} \; .
\ee
For the considered uniform system, the natural single-particle states, 
which are the eigenfunctions of the single-particle density matrix, are 
the plane waves
$$
\vp_k(\vr) = \frac{1}{\sqrt{V}}\; e^{i\vk\cdot\vr} \; .
$$
Introducing the notation for the radius of the system,
\be
\label{24}
R\equiv \left ( \frac{3N}{4\pi\rho}\right )^{1/3} \; ,
\ee
for the matrix elements of $\hat\rho_0=[\rho_0(\vr,\vr')]$ we have
$$
(\vp_k,\hat\rho_0\vp_k) = 2\pi\; \frac{\rho_0 R}{k k_0} \left [
\frac{\sin(k-k_0)R}{(k-k_0)R} \; - \; 
\frac{\sin(k+k_0)R}{(k+k_0)R}\right ] \; .
$$

In particular, if $k_0\ra 0$, then 
$$
\lim_{k_0\ra 0} \left ( \vp_k,\hat\rho_0\vp_k\right ) =  4\pi\;
\frac{\rho_0 R}{k^2} \left ( \frac{\sin{k R}}{k R} \; -
\cos kR \right ) \; .
$$
From here, the largest eigenvalue of the single-particle density matrix is
$$
\sup_k \; \lim_{k_0\ra 0} \left (\vp_k,\hat\rho_0\vp_k\right )
= N_0 \; ,
$$
which results in the order index (18).

However, for $k_0\neq 0$, we find
$$
\sup_k\left (\vp_k,\hat\rho_0\vp_k\right ) = 2\pi \;
\frac{\rho_0 R}{k_0^2}\left [ 1 - \;
\frac{\sin(2k_0 R)}{2k_0 R}\right ] \; .
$$
For large systems, such that $k_0R\gg 1$, we obtain
$$
||\hat\rho_1|| \simeq 2\pi \; \frac{\rho_0}{k_0^2}\left (
\frac{3N}{4\pi\rho}\right )^{1/3} \; ,
$$
which requires that the number of particles be large,
$$
N\gg \frac{4\pi\rho}{3k_0^3} \; .
$$
Thus, for the order index of the single-particle matrix, we obtain 
$\om_1(\psi)=1/3$. In the general case, exploiting the relation
$$
||\hat\rho_p|| \sim ||\hat\rho_1||^p \; ,
$$
valid for asymptotically large $N$, we have the order indices (17) as
\be
\label{25}
\om_p(\psi) = \frac{1}{3} \; .
\ee
This value of the order indices corresponds to {\it mid-range order}.

If, instead of the spectral norm (4), one uses the trace norm (5), then 
the order indices for reduced density matrices are different. To calculate 
such indices, we notice that
$$
{\rm Tr}\; \hat\rho_0^2 = 2\pi V \; \frac{\rho_0^2 R}{k_0^2} \left [
1 - \; \frac{\sin(2k_0 R)}{2k_0 R}\right ] \; .
$$
For the standard condensate with $k_0=0$, we have ${\rm Tr}\hat\rho_0^2
= N_0^2$, from where $\om_1'(\psi)=1$, which coincides with $\om_1(\psi)=1$. 
But if $k_0\neq 0$ and the system is large, such that $k_0 R\gg 1$, then 
we find
$$
||\hat\rho_1||'\simeq \sqrt{\frac{8}{3}}\; \pi\; \frac{\rho_0}{k_0}
\left ( \frac{3N}{4\pi\rho}\right )^{2/3} \; .
$$
Finally, we obtain
\be
\label{26}
\om_p'(\psi) = \frac{2}{3} \; ,
\ee
which differs from the index (25).

\subsection{Two-dimensional case}

The condensate density matrix (21) is 
\be
\label{27}
\rho_0(\vr,\vr') = \rho_0 J_0(k_0|\vr -\vr'|) \; ,
\ee
where
$$
J_n(z) \equiv \frac{(-1)^n}{\pi} \;
\int_0^\pi e^{iz\cos\vp}\cos(n\vp) \; d\vp
$$
is the first-type Bessel function, with $n=0,1,2,\ldots$ Using the 
relation
$$
J_n\left ( ze^{i\pi m}\right ) = e^{i\pi mn} J_n(z) \; ,
$$
in which $m=0,\pm 1,\pm 2,\ldots$, and the integral
$$
\int xJ_n(ax) J_n(bx)\; dx = \frac{x}{a^2-b^2} \left [
aJ_{n+1}(ax)J_n(bx) - b J_n(ax) J_{n+1}(bx) \right ] \; ,
$$
we find
$$
(\vp_k,\hat\rho_0\vp_k) = \frac{2\pi\rho_0 R}{k^2-k_0^2}\left [
kJ_1(kR) J_0(k_0R) - k_0 J_0(kR) J_1(k_0 R)\right ] \; ,
$$
where
\be
\label{28}
R \equiv \left ( \frac{N}{\pi\rho}\right )^{1/2}
\ee
is the system radius.

When $k_0\ra 0$, then
$$
\lim_{k_0\ra 0} (\vp_k,\hat\rho_0\vp_k) = 2\pi\; 
\frac{\rho_0 R}{k} \; J_1(k R) \; .
$$
Because of the asymptotic equation
$$
J_n(z) \simeq \frac{1}{n!}\left ( 
\frac{z}{2}\right )^n \qquad (z\ra 0) \; ,
$$
it follows that
$$
\sup_k 2\pi\; \frac{\rho_0 R}{k}\; J_1 (kR) = 
\pi R^2\rho_0 = N_0 \; .
$$
And we again return to the indices (18).

However, for the finite-momentum condensate, with $k_0\neq 0$, we 
have
$$
\sup_k \left (\vp_k,\hat\rho_0\vp_k\right ) = \pi\rho_0 R^2\left [ J_0^2(k_0 R)
 - \; \frac{J_0(k_0R)J_1(k_0R)}{k_0R} +
J_1^2(k_0R) \right ] \; ,
$$
where the equations
$$
z\; \frac{dJ_n(z)}{dz} =  nJ_n(z) - z J_{n+1}(z) = 
z J_{n-1}(z) - n J_n(z)
$$
have been used. For a large system, such that $k_0R\gg 1$, we invoke the 
asymptotic form
$$
J_n(z) \simeq \sqrt{\frac{2}{\pi z}}\; \cos\left ( z -\; 
\frac{\pi}{2}\; n - \; \frac{\pi}{4}\right ) \qquad 
(z \ra \infty) \; .
$$
Then, for a system of many particles, with
$$
N\gg \frac{\pi\rho}{k_0^2} \; ,
$$
we get
$$
||\hat\rho_1|| \simeq 4\; \frac{\rho_0}{k_0}\left (
\frac{N}{\pi\rho}\right )^{1/2} \cos^2(k_0 R) \; .
$$
For the order indices we obtain
\be
\label{29}
\om_p(\psi) = \frac{1}{2} \; .
\ee

If calculating the order indices, one opts for the trace norm (5), then 
the resulting $\om_p'(\psi)$ differs from the indices (29). Thus, using 
the relations
$$
J_{-n}(z) = (-1)^n J_n(z) \; ,
$$
$$
\int xJ_n^2(ax)\; dx = \frac{x^2}{2}\left [ J_n^2(ax) -
J_{n-1}(ax)J_{n+1}(ax)\right ] \; ,
$$
we have
$$
{\rm Tr}\; \hat\rho_0^2 =\left ( \rho_0\pi R^2\right )
\left [ J_0^2(k_0 R) + J_1^2(k_0 R)\right ] \; .
$$

The indices $\om_p'(\psi)$ coincide with $\om_p(\psi)=1$ only if $k_0=0$, 
since then ${\rm Tr}\hat\rho_0^2=N_0^2$. But for $k_0\neq 0$, again 
keeping in mind a large system, with $k_0R\gg 1$, we find
$$
||\hat\rho_1||' \simeq 2\rho_0\sqrt{\frac{\pi}{k_0}}\left (
\frac{N}{\pi\rho}\right )^{3/4}\; |\cos(k_0 R)| \; .
$$
As a result, we obtain
\be
\label{30}
\om_p'(\psi) = \frac{3}{4} \; ,
\ee
which is different from the indices (29).

\subsection{One-dimensional case}

The condensate density matrix (21) takes the form
\be
\label{31}
\rho_0(x,x') = \rho_0\cos k_0(x-x') \; ,
\ee
where $x\in(-\infty,+\infty)$. With the system length
\be
\label{32}
L\equiv \frac{N}{\rho} \; ,
\ee
we have
$$
(\vp_k,\hat\rho_0\vp_k) = \rho_0 L\left [
\frac{\sin\left (\frac{k-k_0}{2}\; L\right )}{(k-k_0)L} +
\frac{\sin\left (\frac{k+k_0}{2}\; L\right )}{(k+k_0)L}
\right ] \; .
$$

If $k_0=0$, then again we get $\om_p(\psi)=1$, since
$$
\lim_{k_0\ra 0} (\vp_k,\hat\rho_0\vp_k) = 2\; \frac{\rho_0}{k} \;
\sin \left ( \frac{kL}{2}\right )
$$
and the supremum over $k$ for the above right-hand side gives $N_0$.

The situation is rather interesting for $k_0\neq 0$. Then
$$
\sup_k\left (\vp_k,\hat\rho_0\vp_k\right ) = \frac{\rho_0 L}{2}
\left [ 1 + \frac{\sin(k_0 L)}{k_0 L} \right ] \; .
$$
For a large system, with $k_0 L\gg 1$, and, respectively, with a large 
number of particles, such that
$$
N\gg \frac{\rho}{k_0} \; ,
$$
we find
$$
||\hat\rho_1|| \simeq \frac{\rho_0}{2\rho}\; N \; .
$$
Finally, we obtain
\be
\label{33}
\om_p(\psi) =  1 \; .
\ee

When employing the trace norm (5), and using the equality 
$$
{\rm Tr}\hat\rho_0^2 = \frac{1}{2}\left (\rho_0 L\right )^2 \left (
1 + \frac{\sin k_0 L}{k_0 L}\right ) \; ,
$$
we get for a large system
$$
||\hat\rho_1||' \simeq \frac{\rho_0}{\sqrt{2}\; \rho}\; N \; .
$$
This yields

\be
\label{34}
\om_p'(\psi) =  1 \; .
\ee
Hence, in the case of one-dimensional systems, the order indices (33), 
based on the spectral norm, coincide with the indices (34), based on the 
trace norm of reduced density matrices.

The order indices, defined through the spectral norm, or respectively 
through the trace norm, permit us to generalize the consideration to an 
arbitrary $d$-dimensional case resulting in
\be
\label{35}
\om_p(\psi) = \frac{1}{d}\; , \qquad \om_p'(\psi) = 
\frac{d+1}{2d} \; .
\ee
This suggests a relation
$$
\om_p'(\psi) = \frac{1+\om_p(\psi)}{2}
$$ 
between the order indices obtained for different norms.

Let us emphasize that the order indices provide us with a quantitative 
measure for the degree of order occurring in a statistical system. This 
measure makes it possible to unambiguously distinguish between the 
degrees of order of quite similar systems. For example, let us compare 
two three-dimensional systems, one with the condensate density matrix (23)
and another with that matrix having the same algebraic decay, but without 
oscillations due to sine, that is when $\rho_0(\vr,\vr')\sim
\rho_0/k_0|\vr-\vr'|$. For the latter case, we find $\om_p(\psi)=2/3$ 
instead of the index (25), which tells us that in the latter system 
the order is higher than in the system with the condensate matrix (23).

It is important that the order indices allow the description and 
classification of various types of mid-range order, as is illustrated 
in this section, when the order parameters are zero. Note that a Bose 
system with finite-momentum condensate [10--12], possessing mid-range 
order, satisfies the Cummings-Hyland-Rowlands relation [13].

\section{Condensation in confined systems}

The recent developments of Bose-Einstein condensation of trapped atoms 
(see reviews [14--16]) makes it important to consider how the order 
indices can be defined for finite systems. Stricktly speaking, in finite 
systems sharp phase transitions cannot occur. A rigorous definition of 
a genuine phase transition requires the usage of the thermodynamic 
limit (16). But the number of atoms in a trap, although large, however 
is always finite. Therefore the Bose-Einstein condensation in a trap is 
rather a gradual crossover than a sharp transition. The introduced 
matrix order indices do not necessarily require the employment of the 
thermodynamic limit, though their calculation is facilitated when the 
number of particles is large.

Bose-Einstein condensation can be treated as the appearance of a 
macroscopic fraction of atoms being in a coherent state. A coherent 
state, in general, is defined up to a phase factor with a random 
phase [17]. If this phase is made fixed, then the gauge symmetry 
in the system is broken. The breaking of gauge symmetry is usually 
done by means of the Bogolubov prescription [18] presenting the 
field operator as $\psi=\psi_0+\tilde\psi$, where $\psi_0$ is a 
classical quantity and $\tilde\psi$ is the field operator of 
noncondensed particles. In that case, the statistical average 
$<\psi>=\psi_0$ is nonzero and might be considered as an order 
parameter. The gauge symmetry breaking implies that the number of 
particles in the system is not conserved. In reality, the number 
of particles in a system, even such as a trap, can be preserved 
with a rather good accuracy. Consequently, it would be logical to 
describe Bose-Einstein condensation without breaking gauge symmetry 
[19--24]. The suggested schemes, preserving particle conservation, 
are essentially more cumbersome then the techniques based on gauge 
symmetry breaking. This is why the latter techniques became more 
popular. However this way also contains intrinsic difficulties related 
with the appearance of divergences in low-order approximations. To 
compensate these divergences, one needs to invoke higher-order 
approximations [23], which again complicates the consideration. Here 
we advance a novel method of treating a system with Bose condensate, 
without breaking gauge symmetry. The method is sufficiently simple 
and general, and can be applied to trapped as well as untrapped 
atoms interacting through integrable potentials.

The pivotal idea of the method is the usage of {\it random-phase 
coherent states} [17]. When a fraction of atoms is in such a state, 
the field operator can be presented as the sum
\be
\label{36}
\psi(\vr,t) =\eta_\al(\vr,t) +\tilde\psi(\vr,t) \; ,
\ee
in which
\be
\label{37}
\eta_\al(\vr,t) = \eta(\vr,t) e^{i\al}
\ee
is a {\it random-phase coherent field} [17] with $\al\in[0,2\pi]$ being a 
random phase, and $\tilde\psi$ corresponds to incoherent particles.

The average of an operator $\hat A$ is defined as
\be
\label{38}
<\hat A> \; \equiv \frac{1}{2\pi} \int_0^{2\pi} {\rm Tr}\hat\rho
\hat A\; d\al \; ,
\ee
including statistical averaging, with a statistical operator $\hat\rho$, 
and averaging over the random phases. From this definition, it follows that
\be
\label{39}
<\psi>\; = \; <\eta_\al>\; = \; <\tilde\psi>\; = 0 \; ,
\ee
hence the gauge symmetry is not broken. The {\it conservation of gauge 
symmetry} is what makes the principal distinction between the 
presentation (36) and the Bogolubov prescription [18], where the 
gauge symmetry is broken. The coherent and incoherent terms in the 
sum (36) are subject to the normalization conditions
$$
\int <\psi^\dgr(\vr,t)\psi(\vr,t)>\; d\vr = N\; , \qquad
\int |\eta(\vr,t)|^2 d\vr = N_0 \; ,
$$
\be
\label{40}
\int < \tilde\psi^\dgr(\vr,t)\tilde\psi(\vr,t)> \; d\vr
= \tilde N\; ,
\ee
where $N_0$ is the number of coherent atoms, $\tilde N$ is the number of 
incoherent atoms, and the total number of atoms is
\be
\label{41}
N = N_0 +\tilde N \; .
\ee
The terms in the right-hand side of Eq. (36) are mutually orthogonal 
on the average, which means that
$$
<\eta_\al(\vr,t)\tilde\psi(\vr',t)>\; = 0 \; .
$$

For the standard Hamiltonian with a two-particle interaction potential 
$\Phi(\vr)=\Phi(-\vr)$, the Heisenberg equation for the field operator 
reads
\be
\label{42}
i\hbar\; \frac{\prt\psi}{\prt t} =  H(\psi)\psi \; ,
\ee
where $\psi=\psi(\vr,t)$ and
\be
\label{43}
H(\psi) = -\; \frac{\hbar^2{\bf\nabla}^2}{2m_0} +
U_{ext} + U_{int} - \mu \; ,
\ee
with $m_0$ being particle mass and $\mu$, chemical potential. The external 
potential $U_{ext}(\vr,t)$ describes all external fields, including the 
confining potential and possible external perturbations. The interaction 
potential 
\be
\label{44}
U_{int}(\vr,t) = \int \Phi(\vr-\vr')\psi^\dgr(\vr',t)\psi(\vr',t)\; d\vr' 
\ee
is formed by particle interactions.

Let us substitute the presentation (36) into the Heisenberg equation (42). 
Then, let us either multiply Eq. (42) by $e^{-i\al}$ or just set $\al=0$, 
after which take the average $<\ldots>$ according to definition (38). 
This results in the equation for the coherent field,
$$
\left ( i\hbar\; \frac{\prt}{\prt t} + 
\frac{\hbar^2{\bf\nabla}^2}{2m_0}\; - U_{ext} +\mu \right ) 
\eta(\vr,t) = 
$$
\be
\label{45}
= \int \Phi(\vr-\vr') \left [ |\eta(\vr',t)|^2 \eta(\vr,t)
+ \tilde\rho(\vr',t)\eta(\vr,t) + \tilde\rho_1(\vr,\vr',t)\eta(\vr',t)
\right ]\; d\vr' \; ,
\ee
in which the density of incoherent particles
\be
\label{46}
\tilde\rho(\vr,t) \equiv \tilde\rho_1(\vr,\vr,t)
\ee
is the diagonal element of the incoherent density matrix
\be
\label{47}
\tilde\rho_1(\vr,\vr',t) \equiv \; 
<\tilde\psi^\dgr(\vr',t)\tilde\psi(\vr,t)> \; .
\ee
Notice that in the equation (45) for the coherent field no anomalous 
averages arise, which is due to the gauge symmetry conservation.

If, after substituting the presentation (36) into the Heisenberg 
equation (42), we average only over the random phase, then we obtain
the following equation
$$
\left ( i\hbar\; \frac{\prt}{\prt t} + 
\frac{\hbar^2{\bf\nabla}^2}{2m_0} \; - U_{ext} +\mu\right ) 
\tilde\psi(\vr,t) =
$$
\be
\label{48}
=\int \Phi(\vr-\vr') \left [ 
\tilde\psi^\dgr(\vr',t)\tilde\psi(\vr',t)\tilde\psi(\vr,t) +
|\eta(\vr',t)|^2\tilde\psi(\vr,t) +\eta^*(\vr',t)\eta(\vr,t)
\tilde\psi(\vr',t) \right ] \; d\vr'
\ee
for the incoherent term of the field operator. Let us stress that both 
equations (45) as well as (48) are exact.

One often obtains an equation for the condensate wave function, which
is equivalent to Eq. (45), following what one calls the Popov approximation. 
In that way, one, first, breaks the gauge symmetry by means of the Bogolubov 
prescription and then one omits the anomalous averages arising because of 
the broken gauge symmetry. Such a way is clearly not self-consistent. 
Moreover, as is possible to show by direct calculations, in the case of 
broken gauge symmetry, anomalous averages are of the same order, or even 
larger, then normal averages [25,26]. Therefore, after the gauge symmetry 
has been broken, it is not correct to omit anomalous averages.

Given Eq. (48), we may derive an equation for the Green functions of 
incoherent particles. For this purpose, it is convenient to introduce
the following short-hand notation [17]. Denote a function $f(\vr_1,t_1)$ 
as $f(1)$ and the related differential measure as $d(1)\equiv  d\vr_1dt_1$. 
Then the Dirac delta-function is
$$
\dlt(12)\equiv \dlt(\vr_1 -\vr_2) \dlt(t_1-t_2) \; .
$$
Introduce the retarded interaction potential
$$ 
\Phi(12) \equiv \Phi(\vr_1 -\vr_2)\dlt(t_1 -t_2+0) \; .
$$
With this notation, the single-particle and two-particle causal Green 
functions are defined as
\be
\label{49}
G(12)\equiv -i\; <\hat T\; \tilde\psi(1)\tilde\psi^\dgr(2)>\; , \qquad 
G(1234)\equiv -\; <\hat T\; \tilde\psi(1)\tilde\psi(2)
\tilde\psi^\dgr(3)\tilde\psi^\dgr(4) >\; ,
\ee
where $\hat T$ is the chronological operator. These are the Green 
functions at real times, which are directly related to the Matsubara
Green functions at imaginary times [17,27].

From definition (49) and Eq. (48), we immediately obtain the propagator 
equation
$$
\left ( i\hbar\; \frac{\prt}{\prt t_1} + 
\frac{\hbar^2{\bf\nabla}^2_1}{2m_0} - U_{ext} + \mu \right ) G(12) -
$$
\be
\label{50}
- \int \Phi(13)\left [ |\eta(3)|^2 G(12) +\eta^*(3)\eta(1) G(32)
+i G_2(1332)\right ] \; d(3) = \dlt(12) \; .
\ee

It is important to note that all Eqs. (45), (48), and (50) have sense 
only for the integrable interaction potentials, such that
\be
\label{51}
\left | \int_V \Phi(\vr)\; d\vr \right | < \infty \; .
\ee
In the other case, to avoid ultraviolet divergences, one has to set 
$\eta(\vr,t)\equiv 0$, which implies the absence of condensate [17].

The propagator equation (50) can be written in a more compact form if 
one introduces the self-energy $\Sigma(12)$ by means of the relation
$$
\int \Sigma (13) G(23)\; d(3) =
$$
\be
\label{52}
= \int \Phi(13)\left [ |\eta(3)|^2 G(12) +\eta^*(3)\eta(1) G(32) + 
i G_2(1332)\right ] \; d(3) \; .
\ee
From here, recalling the definition of the inverse propagator,
$$
\int G^{-1}(13) G(32) \; d(3) = 
\int G(13) G^{-1}(32) \; d(3) = \dlt(12) \; ,
$$
we have the self-energy
$$
\Sigma (12) = \dlt(12) \int \Phi(13) |\eta(3)|^2 \; d(3) +
$$
\be
\label{53}
+ \Phi(12) \eta(1) \eta^*(2) + i \int \Phi(13) G_2(1334) G^{-1}(42)\;
d (34) \; .
\ee
Then the inverse propagator is
\be
\label{54}
G^{-1}(12) = \left ( i\hbar\; \frac{\prt}{\prt t_1} +
\frac{\hbar^2{\bf\nabla}^2_1}{2m_0}\; - U_{ext} + \mu\right )
\dlt(12) - \Sigma (12) \; ,
\ee
and the propagator equation (50) takes the form
\be
\label{55}
\left ( i\hbar \; \frac{\prt}{\prt t_1} +
\frac{\hbar^2{\bf\nabla}^2_1}{2m_0} - U_{ext} + \mu \right ) G(12) -
\int \Sigma (13) G(32)\; d(3) = \dlt(12) \; .
\ee
The pair of equations (45) and (55), together with the definition (53) 
for the incoherent self-energy, completely describes the Bose system 
with a coherent condensate. These equations, for a given interaction 
potential, are exact.

Dilute gases of trapped atoms are usually characterized by the Fermi 
contact potential
\be
\label{56}
\Phi(\vr) = A \dlt(\vr) \; , \qquad
A\equiv 4\pi \hbar^2 \; \frac{a_s}{m_0} \; ,
\ee
where $a_s$ is a scattering length.

In that case, the coherent-field equation (45) simplifies to
\be
\label{57}
i\hbar\; \frac{\prt\eta}{\prt t}  =\left [ -\; 
\frac{\hbar^2{\bf\nabla}^2}{2m_0} + U_{ext} + A \left ( |\eta|^2 +
2\tilde\rho \right ) - \mu \right ] \eta \; ,
\ee
where $\eta=\eta(\vr,t)$ and $\tilde\rho=\tilde\rho(\vr,t)$. Equation (48) 
for the incoherent operator becomes
\be
\label{58}
i\hbar\; \frac{\prt\tilde\psi}{\prt t}  =\left [ -\; 
\frac{\hbar^2{\bf\nabla}^2}{2m_0} + U_{ext} + A \left ( \tilde\psi^\dgr
\tilde\psi + 2|\eta|^2 \right ) - \mu \right ] \tilde\psi \; .
\ee
The self-energy (53) transforms to
\be
\label{59}
\Sigma (12) =  A\left [ 2\dlt(12) |\eta(1)|^2 + 
i \int G_2(1113) G^{-1}(32) \; d (3)\right ] \; ,
\ee
which enters the propagator equation (55).

To solve the propagator equation, one may follow one of the known
iterative procedures [17,28] starting e.g. with the Hartree-Fock 
approximation, when
$$
G_2(1234) = G(14)G(23) + G(13)G(24) \; .
$$
Then the self-energy (59) is
\be
\label{60}
\Sigma (12) = 2 A\dlt(12)\left [ |\eta(1)|^2 + \tilde\rho(1)\right ] \; .
\ee

If the external potential does not depend on time, $U_{ext}=U(\vr)$, 
and the system is in equilibrium, then the coherent field may be 
presented as
\be
\label{61}
\eta(\vr,t) = \sqrt{N_0}\; \vp(\vr)\; \exp\left ( - \;
\frac{i}{\hbar}\; Et\right ) \; ,
\ee
with the function $\vp(\vr)$ normalized to unity, $(\vp,\vp)=1$. 
In this case, Eq. (57) reduces to the stationary equation
\be
\label{62}
\hat H(\vp)\vp(\vr) = E\vp(\vr) \; ,
\ee
with the effective Hamiltonian
\be
\label{63}
\hat H(\vp) \equiv -\; \frac{\hbar^2{\bf\nabla}^2}{2m_0} +
U + A\left ( N_0 |\vp|^2 + 2\tilde\rho\right ) - \mu \; .
\ee
Note that this Hamiltonian depends on thermodynamic parameters, such as
temperature, through the incoherent density $\tilde\rho=\tilde\rho(\vr)$.

In the presence of a confining potential, the spectrum of the 
eigenproblem (62) is discrete [29]. For a system in equilibrium, the
condensate energy $E$ corresponds to the lowest energy level of the 
eigenproblem (62). For a nonequilibrium system, one can excite higher 
nonlinear coherent modes corresponding to higher states of Eq. (62), 
thus, creating nonground-state coherent condensates [29--33].

For an equilibrium system, the single-particle density matrix
$$
\rho_1(\vr,\vr') = \; <\psi^\dgr(\vr')\psi(\vr)>
$$
is the sum
\be
\label{64}
\rho_1(\vr,\vr') = \rho_0(\vr,\vr') + \tilde\rho_1(\vr,\vr')
\ee
of the coherent term
$$
\rho_0(\vr,\vr') \equiv N_0\vp_0(\vr)\vp_0^*(\vr') \; ,
$$
where $\vp_0(\vr)$ is the ground state of the eigenproblem (62), and 
of the incoherent density matrix (47). The coherent density matrix 
$\hat\rho_0=[\rho_0(\vr,\vr')]$ has the properties
$$
\hat\rho^p = N_0^{p-1}\hat\rho_0 \; , \qquad
{\rm Tr}\; \hat\rho_0^p = N_0^p \; .
$$
The incoherent density matrix can be presented as an expansion
$$
\tilde\rho_1(\vr,\vr') = \sum_k n_k \vp_k(\vr) \vp_k^*(\vr')
$$
over natural wave functions [3] corresponding to incoherent states.

When the coherent condensate is absent, $N_0\equiv 0$, then the spectral 
norm of $\hat\rho_1$ is $||\hat\rho_1||=\sup_k n_k$, which is finite. The
trace norm gives $||\hat\rho_1||'=\sqrt{\sum_k n_k^2}$, which is of
order $N^{1/2}$. The related order indices are $\om_p(\psi)=0$ and
$\om_p'(\psi)=1/2$, respectively. However, as soon as $N_0>\sup_k n_k$,
then for a large system with $N\gg 1$, one has
\be
\label{65}
\om_p(\psi) = \om_p'(\psi) = \frac{\ln N_0}{\ln N} \; .
\ee
In the thermodynamic limit (16), we return to the order indices (18).
But for a finite system the order indices (65) are, in general, functions
of thermodynamic variables, such as temperature and density. When $N_0$ 
changes from $1$ to $N$, the order indices increase from $0$ to $1$,
which corresponds to a gradual crossover.

Recall that the present consideration has been based on the decomposition
(36) for the field operator onto its coherent and incoherent parts.
This decomposition is valid only when the interaction potential is
integrable, satisfying condition (51). The Fermi contact potential (56)
is integrable. This potential describes pair interactions in dilute 
gases, whose mean interatomic distance is much larger than the magnitude 
of the scattering length $|a_s|$. For dense gases and liquids, interaction
potentials often contain hard cores and are not integrable. In such cases, 
the decomposition (36) is not appropriate and one has to invoke other 
calculational techniques explicitly taking account of interatomic 
correlations (see e.g. [34--36] and references in review [16]).

\section{Magnetic phase transitions}

The matrix indices (17), defined for reduced density matrices, 
are convenient for characterizing such phase transitions as Bose 
condensation and superconductivity. The question that one could pose 
is whether the same indices (17) would be suitable for describing 
magnetic phase transitions. And also, since superconductivity and 
magnetic order can coexist [37,39], could the indices (17) characterize 
such a coexistence of orders? In this section, we show that the order
indices (17), based on reduced density matrices (18), are not suitable
for describing magnetic order. To describe the latter, it is necessary
to return to the general definition (1) of the matrix order indices 
and to introduce the matrices composed of spin operators.

Consider, first, the indices $\om_p(\psi)$. Let the field operator be
a spinor $\psi=[\psi_\sgm]$, with $\sgm=\uparrow,\downarrow$ denoting
spin up or down. Keeping in mind a solid with a crystalline lattice, we 
enumerate the lattice sites by the index $i=1,2,\ldots,N$. Wave 
functions, localized at the related sites, are called the localized 
orbitals, such as Wannier functions. For simplicity, we shall deal with
the single-zone case. Let $\{\vp_i(\vr)\}$ be a set of localized orbitals
that are orthonormalized, $(\vp_i,\vp_j)=\dlt_{ij}$, and asymptotically
complete, which means that
\be
\label{66}
\sum_{i=1}^N \vp_i(\vr)\vp_i^*(\vr') = \dlt_N(\vr-\vr') \; ,
\ee
with the right-hand side asymptotically, as $N\ra\infty$, coinciding with
the $\dlt$-function,
\be
\label{67}
\lim_{N\ra\infty} \dlt_N(\vr) =\dlt(\vr) \; .
\ee

The field operator $\psi_\sgm(\vr)$ can be expanded over the localized 
orbitals as
\be
\label{68}
\psi_\sgm(\vr) = \sum_{i=1}^N c_{i\sgm}\vp_i(\vr) \; .
\ee
Both operators $\psi_\sgm(\vr)$ as well as $c_{i\sgm}$ satisfy the Fermi 
commutation relations. To stress that the particles are well localized,
the following {\it localization conditions} are imposed on the operators
$c_{i\sgm}$:
\be
\label{69}
c_{i\sgm}^\dgr c_{j\sgm'}  =\dlt_{ij} c_{i\sgm}^\dgr c_{i\sgm'} \; ,
\qquad c_{i\sgm}^\dgr c_{j\sgm'}^\dgr c_{f\sgm'} c_{g\sgm} =
\dlt_{ig} \dlt_{jf} c_{i\sgm}^\dgr c_{j\sgm'}^\dgr c_{j\sgm'} c_{i\sgm} 
+ \dlt_{if}\dlt_{jg} c^\dgr_{i\sgm} c^\dgr_{j\sgm'}
c_{i\sgm'} c_{j\sgm} \; .
\ee
Limiting the number of particles at each site by one, we have the
{\it unipolarity conditions}
\be
\label{70}
\sum_\sgm c_{i\sgm}^\dgr c_{i\sgm} = 1 \; , \qquad 
c_{i\sgm} c_{i\sgm'}  = 0 \; .
\ee
From the particle operators $c_{i\sgm}$, one can pass to spin operators
$S_i^\al$ by means of the transformations
$$
c^\dgr_{i\uparrow} c_{i\uparrow} = \frac{1}{2} + S_i^z \; , \qquad
c^\dgr_{i\downarrow} c_{i\downarrow} = \frac{1}{2} - S_i^z \; , 
$$
$$
c^\dgr_{i\uparrow} c_{i\downarrow} = S_i^+ \equiv S_i^x + i S_i^y \; ,
\qquad
c_{i\downarrow}^\dgr c_{i\uparrow} = S_i^- \equiv S_i^x - i S_i^y \; .
$$

Now, we construct the reduced density matrices (13). The single-particle 
density matrix is
\be
\label{71}
\rho_1(\vr,\vr') = \dlt_N (\vr -\vr') \; ,
\ee
where $\dlt_N(\vr)$ is the asymptotic $\dlt$-function, appearing in 
Eq. (66). Note that, because of ${\rm Tr}\hat\rho_1=N$, one has 
$\dlt_N(0)\equiv N/V$. The spectral and trace norms, respectively, give
$$
||\hat\rho_1|| \simeq 1\; , \qquad ||\hat\rho_1||'\simeq \sqrt{N}
\qquad (N\ra \infty) \; .
$$
From here
\be
\label{72}
\om_1(\psi) = 0 \; , \qquad \om_1'(\psi) = \frac{1}{2} \; ,
\ee
independently from the existence or absence of magnetic order.

For the two-particle reduced density matrix, we have
$$
\rho_2(\vr_1,\vr_2,\vr_1',\vr_2') = \rho_1(\vr_1,\vr_1')
\rho_1(\vr_2,\vr_2') - \; \frac{1}{2}\; \rho_1(\vr_1,\vr_2')
\rho_1(\vr_2,\vr_1') -
$$
\be
\label{73}
 - 2 \sum_{i\neq j}^N < {\bf S}_i \cdot {\bf S}_j> \; 
\vp_i(\vr_1)\vp_j(\vr_2) \vp^*_j(\vr_1') \vp_i^*(\vr_2') \; .
\ee
To simplify the consideration, let the particle interactions be of 
long-range type, for which the asymptotic, as $N\ra\infty$, decoupling
$$
<{\bf S}_i \cdot{\bf S}_j> \; \cong \; < {\bf S}_i> <{\bf S}_j>
\qquad (i\neq j)
$$
is valid. And also let the lattice be ideal, when, because of the 
translation invariance, one has
$$
< {\bf S}_i>\; = \;  <{\bf S}> \; , \qquad {\bf S} \equiv
\frac{1}{N} \sum_{i=1}^N {\bf S}_i \; .
$$
Then the two-particle matrix (73) becomes 
\be
\label{74}
\rho_2(\vr_1,\vr_2,\vr_1',\vr_2') = 
\rho_1(\vr_1,\vr_1')\rho_1(\vr_2,\vr_2') - \left ( \frac{1}{2} +
2<{\bf S}>^2 \right ) \rho_1(\vr_1,\vr_2') \rho_1(\vr_2,\vr_1') \; .
\ee
The eigenfunctions of this matrix are
$$
\vp_{ij}(\vr,\vr') = \frac{1}{\sqrt{2}} \left [ \vp_i(\vr)\vp_j(\vr') -
\vp_i(\vr')\vp_j(\vr)\right ] \; ,
$$
with $\vp_i(\vr)$ being the localized orbitals. For a large system, 
we get
$$
||\hat\rho_2|| \simeq \frac{3}{2}  + 2<{\bf S}>^2 \; , \qquad
||\hat\rho_2||' \sim N \; ,
$$
as $N\ra\infty$. Thus, in the thermodynamic limit, we find
\be
\label{75}
\om_p(\psi) = 0 \; , \qquad \om_p'(\psi) = \frac{1}{2} \; .
\ee
This shows that, although the norms of the reduced density matrices do 
change under the arising magnetic order, that change is negligible in the
thermodynamic limit. Hence, the indices $\om_p(\psi)$ are not suitable 
for describing magnetic phase transitions.

This problem can easily be resolved by going back to the more general 
definition of the matrix indices (10). Being interested in magnetic order, 
it is reasonable to work with correlation functions composed of spin 
operators. For instance, the two-spin correlator can be written as
\be
\label{76}
D_{ij}(S) \equiv \; <{\bf S}_j \cdot{\bf S}_i> \; .
\ee
In general, the $2p$-spin correlator is
\be
\label{77}
D_{i_1\ldots i_p j_1\ldots j_p} (S) \equiv <
{\bf S}_{j_p}\cdot\cdot\cdot ({\bf S}_{j_1}\cdot{\bf S}_{i_1})
\cdot\cdot\cdot {\bf S}_{i_p} > \; .
\ee
The related {\it spin-correlation matrices} $\hat D_p(S)$ are the 
matrices with the elements (77). For example, $\hat D_1(S)=[D_{ij}(S)]$
is an $N\times N$ matrix with the elements (76).

For a system defined on an ideal lattice, the natural wave functions are
\be
\label{78}
\vp_k({\bf a}_i)  =\frac{1}{\sqrt{N}}\; e^{i{\bf k}\cdot{\bf a}_i} \; ,
\ee
with ${\bf a}_i$ being a lattice vector. These functions form an 
orthonormalized basis, for which
$$
\sum_i \vp_k^*({\bf a}_i) \vp_{k'}({\bf a}_i) = \dlt_{kk'} \; , \qquad
\sum_k \vp_k({\bf a}_i) \vp_k^*({\bf a}_j) = \dlt_{ij} \; ,
$$

Calculating the spectral norm of $\hat D_1(S)$, we find
$$
||\hat D_1(S)|| = \sup_k \left [ S(S+1) + 
\sum_{j=1}^{N-1} <{\bf S}_j \cdot{\bf S}_i> \; e^{i \vk \cdot{\bf a}_j}
\right ] \; .
$$
The trace norm yields
$$
||\hat D_1(S)||' =\left ( N \sum_{j=1}^{N-1} < {\bf S}_j \cdot
{\bf S}_0 > \right )^{1/2} \; .
$$
And for the trace, we have
$$
{\rm Tr}\; \hat D_1(S) = S(S+1) N \; .
$$
When magnetic order is absent, i.e. $<{\bf S}> = 0$, then for a 
large system, with $N\gg 1$,
$$
||\hat D_1(S)|| \sim 1 \; , \qquad ||\hat D_1(S)||' \sim N^{1/2} \; .
$$
Hence, generalizing this, we get
\be
\label{79}
\om_p(S) = 0 \; , \qquad \om_p'(S) = \frac{1}{2} \qquad
(<{\bf S}> \; = 0 ) \; .
\ee
But as soon as there appear magnetic order, then
$$
||\hat D_1(S)|| \simeq N \; <{\bf S}>^2 \; , \qquad
||\hat D_1(S)||'\simeq N|<{\bf S}>| \; ,
$$
as $N\ra\infty$. In this way, we come to the order indices
\be
\label{80}
\om_p(S) =\om_p'(S) = 1 \qquad (<{\bf S}>\; \neq 0) \; .
\ee
Comparing Eqs. (79) and (80), we see that the order indices $\om_p(S)$ 
perfectly distinguish between the existence and absence of magnetic 
order. The description of the latter by the indices $\om_p(S)$ is
analogous to the characterization of Bose condensation by the indices 
$\om_p(\psi)$. This is not surprising, since the arising magnetic order 
can be interpreted as condensation of magnons [39].

\section{Transitions with diagonal order}

The eigenvalues of the reduced density matrices $\hat\rho_p$ are known
to be sensitive to the existence of the off-diagonal long-range order
[3--6]. This is why the indices $\om_p(\psi)$ are so useful for 
characterizing phase transitions with arising off-diagonal order, such as 
superconductivity and Bose condensation. Similarly, the indices $\om_p(S)$
are convenient for describing magnetic transitions. These indices, however, 
are not sensitive to the appearance of the so-called diagonal long-range 
order, associated with the properties of the diagonal elements of density
matrices. Among phase transitions with developing diagonal order, we 
may mention the melting-crystallization phase transition, which is so 
well known on earth and, presumably, occurring in neutron stars [40]. 
The mixing-stratification transformation is also of this kind.

The difference between phase transitions with off-diagonal long-range 
order and diagonal long-range order is in the following. In the former 
case, the largest eigenvalue of one of the reduced density matrices, and respectively the related spectral norm, change from being of order one to
the order of $N^p$. Such a dramatic variation of the norm, due to the 
arising off-diagonal order, immediately results in the sharp change of 
the order index $\om_p(\psi)$. But so drastic increase of eigenvalues of 
the reduced density matrices does not happen under the appearing diagonal 
order, when all eigenvalues of $\hat\rho_p$ remain of order one. Thus, for
both crystalline and liquid states
$$
||\hat\rho_p|| \sim 1 \; , \qquad ||\hat\rho_p||'\sim N^{1/2} \; .
$$
Therefore the indices
\be
\label{81}
\om_p(\psi)= 0 \; , \qquad \om_p'(\psi) = \frac{1}{2}
\ee
do not notice that the diagonal order has appeared.

Fortunately, with the general definition of the matrix order indices 
(1) or (10), we are not obliged to deal exclusively with the density 
matrices (12). It is always possible to define such a matrix whose 
order index would be sensitive to any kind of arising order.

As an illustration, let us consider the crystallization transition. 
Introduce the {\it density-difference operator}
\be
\label{82}
\Dlt(\vr) \equiv \psi^\dgr(\vr)\psi(\vr) - \rho \; ,
\ee
with $\rho\equiv N/V$ being the average density. Statistical averaging 
of the operator (82) gives
$$
<\Dlt(\vr)>\; = \rho(\vr) - \rho \; , \qquad \rho(\vr) \equiv \;
<\psi^\dgr(\vr)\psi(\vr)> \; .
$$
Define the matrix $\hat D_1(\Dlt)$ whose elements are 
\be
\label{83}
D_1(\Dlt,\vr,\vr') = \; < \Dlt(\vr')\Dlt(\vr)> \; .
\ee
Similarly, we may construct a matrix $\hat D_p(\Dlt)$ of any order 
$p=1,2,\ldots$, as is explained in Eq. (6). The elements of 
$\hat D_p(\Dlt)$ satisfy the correlation weakening condition, e.g.
$$
<\Dlt(\vr)\Dlt(\vr') > \; \simeq \; <\Dlt(\vr)>\; <\Dlt(\vr')>
\qquad (|\vr -\vr'|\ra \infty) \; .
$$

In the case of liquid, with uniform density, we have
$$
<\Dlt(\vr)> \; = 0 \; , \qquad \rho(\vr) = \rho \; .
$$
Eigenvalues of $\hat D_1(\Dlt)$ are of order $1$, which yields
$$
||\hat D_1(\Dlt)|| \sim 1 \; , \qquad ||\hat D_1(\Dlt)||'\sim
N^{1/2} \; .
$$
Following this way, we find
\be
\label{84}
\om_p(\Dlt) = 0 \; , \qquad \om_p'(\Dlt) = \frac{1}{2} \; ,
\ee
which signifies the absence of order.

The occurrence of crystalline order is marked by a well defined 
crystalline lattice spanned by the lattice vectors ${\bf a}_i$, with
$i=1,2,\ldots,N$. Introduce the function $D_1(\Dlt, {\bf a}_i,{\bf a}_j)$
as in Eq. (83) and the related matrix 
$\hat D_1(\Dlt)= [D_1(\Dlt,{\bf a}_i,{\bf a}_j)]$, which is an $N\times N$
matrix. The natural wave functions for a lattice system are presented 
in Eq. (78). Employing the spectral norm, we get
$$
||\hat D_1(\Dlt)|| \simeq N [ \rho({\bf a}) - \rho ]^2 \qquad
(N\ra \infty) \; ,
$$
where ${\bf a}$ is any of ${\bf a}_i$, since for an ideal lattice, 
$\rho({\bf a}_i) =\rho({\bf a}_j)$. The same expression follows for the 
trace norm $||\hat D_1(\Dlt)||'$. Finally, we obtain
\be
\label{85}
\om_p(\Dlt) = \om_p'(\Dlt) = 1 \; .
\ee
As is seen, the formation of crystalline order leads to the change of 
the matrix indices $\om_p(\Dlt)$ from the values in Eq. (84) to those 
in Eq. (85).

\vskip 2mm

As s brief conclusion, let us stress that the matrix order indices, 
introduced in the present paper, is a new general notion allowing
for the description and classification of arbitrary types of order,
off-diagonal as well as diagonal, long-range and mid-range, occurring
in infinite as well as finite systems, and happening in different spaces 
(real, momentum, spin,$\ldots$). The matrix order indices can be well 
defined when the order parameters do not exist. These indices are suitable 
for describing sharp phase transitions as well as gradual crossovers.

\vskip 5mm

{\bf Acknowledgement}

\vskip 2mm

I am grateful to A.J. Coleman for numerous useful discussions. A Grant of
the S\~ao Paulo State Research Foundation, Brazil, is appreciated.


\newpage

\begin{thebibliography}{99}

\bibitem{1}
D. ter Haar, Rep. Prog. Phys. 24 (1961) 304.

\bibitem{2}
K. Blum, Density Matrix Theory and Applications, Plenum, New York, 1981.

\bibitem{3}
A.J. Coleman, V.I. Yukalov, Reduced Density Matrices, Springer, 
Berlin, 2000.

\bibitem{4}
O. Penrose, Philos. Mag. 42 (1951) 1373.

\bibitem{5}
O. Penrose, L. Onsager, Phys. Rev. 104 (1956) 576.

\bibitem{6}
C.N. Yang, Rev. Mod. Phys. 34 (1962) 694.

\bibitem{7}
A. J. Coleman, V.I. Yukalov, Mod. Phys. Lett. B 5 (1991) 1679.

\bibitem{8}
A.J. Coleman, V.I. Yukalov, Nuovo Cimento B 108 (1993) 1377.

\bibitem{9}
A.J. Coleman, V.I. Yukalov, Int. J. Mod. Phys. B 10 (1996) 3505.

\bibitem{10}
V.I. Yukalov, Theor. Math. Phys. 37 (1978) 1093.

\bibitem{11}
V.I. Yukalov, Physica A 100 (1980) 431.

\bibitem{12}
V.I. Yukalov, Phys. Lett. A 83 (1981) 26.

\bibitem{13}
F.W. Cummings, C.J. Hyland, G. Rowlands, Phys. Lett. A 86 (1981) 370.

\bibitem{14}
A.S. Parkins, D.F. Walls, Phys. Rep. 303 (1998) 1.

\bibitem{15}
F. Dalfovo, S. Giorgini, L.P. Pitaevskii, S. Stringari, Rev. Mod. Phys. 71 (1999) 463.

\bibitem{16}
P.W. Courteille, V.S. Bagnato, V.I. Yukalov, Laser Phys. 11 (2001) 659.

\bibitem{17}
V.I. Yukalov, Statistical Green's Functions, Queen's University, Kingston, 1998.

\bibitem{18}
N.N. Bogolubov, Lectures on Quantum Statistics, Vol. 1, Gordon and Breach, New York, 1967.

\bibitem{19}
M. Girardeau and R. Arnowitt, Phys. Rev. 113 (1959) 755.

\bibitem{20}
C.W. Gardiner, Phys. Rev. A 56 (1997) 1414.

\bibitem{21}
Y. Castin, R. Dum, Phys. Rev. A57 (1998) 3008.

\bibitem{22}
M. Girardeau, Phys. Rev. A 58 (1998) 775.

\bibitem{23}
S.A. Morgan, J. Phys. B 33 (2000) 3847.

\bibitem{24}
A.E. Ruckenstein, Found. Phys. 30 (2000) 2113.

\bibitem{25}
A. Griffin, Phys. Rev. B 22 (1980) 5193.

\bibitem{26}
T. Isoshima, K. Machida, J. Phys. Soc. Jap. 66 (1997) 3502.

\bibitem{27}
G. Cuniberti, E. De Micheli, G.A. Viano, Commun. Math. Phys. 216 (2001) 59.

\bibitem{28}
R. Del Sole, L. Reining, R.W. Godby, Phys. Rev. B 49 (1994) 8024.

\bibitem{29}
V.I. Yukalov, E.P. Yukalova, V.S. Bagnato, Phys. Rev. A 56 (1997) 4845.

\bibitem{30}
V.I. Yukalov, E.P. Yukalova, V.S. Bagnato, Laser Phys. 10 (2000) 26.

\bibitem{31}
V.S. Bagnato, E.P. Yukalova, V.I. Yukalov in: Bose-Einstein Condensates and Atom Lasers, eds. S. Martellucci, A. Chester, A. Aspect, M. Inguscio, Kluwer, New York, 2000, p. 201.

\bibitem{32}
V.I. Yukalov, E.P. Yukalova, V.S. Bagnato, Laser Phys. 11 (2001) 455.

\bibitem{33}
V.I. Yukalov, E.P. Yukalova, V.S. Bagnato, Proc. Int. Soc. Opt. Eng. 4243 (2001) 150.

\bibitem{34}
E. Manousakis, V.K. Pandharipande, Q.N. Usmani, Phys. Rev. B 31 (1985) 7022.

\bibitem{35}
C. Carraro, S.E. Koonin, Phys. Rev. Lett. 65 (1990) 2792.

\bibitem{36}
E. Manousakis, V.K. Pandharipande, Q.N. Usmani, Phys. Rev. B 43 (1991) 13587.

\bibitem{37}
I. Felner, U. Asaf, Y. Levi, O. Millo, Phys. Rev. B 55 (1997) 3374.

\bibitem{38}
I. Felner, U. Asaf, C. Godart, E. Alleno, Physica B 259 (1999) 703.

\bibitem{39}
V.I. Yukalov, A.S. Shumovsky, Lectures on Phase Transitions, World 
Scientific, Singapore, 1990.

\bibitem{40}
N.K. Glendenning, Phys. Rep. 342 (2001) 393.

\end{thebibliography}
\end{document}